\documentclass[lettersize,journal]{IEEEtran}
\usepackage{amsmath,amsfonts}
\usepackage{algorithmic}
\usepackage{array}

\usepackage{textcomp}
\usepackage{stfloats}
\usepackage{url}
\usepackage{verbatim}
\usepackage{graphicx}
\usepackage{cite}
\usepackage{hyperref}
\usepackage{float}
\usepackage{multirow}
\usepackage{makecell}
\usepackage{graphicx} 
\usepackage{booktabs}
\usepackage{epstopdf}
\usepackage[linesnumbered,ruled,vlined]{algorithm2e}
\usepackage[numbers,compress]{natbib}
\usepackage{amsmath,amssymb,amsfonts}

\usepackage[caption=false,font=normalsize,labelfont=sf,textfont=sf]{subfig}

\hypersetup{
    colorlinks=true,
    linkcolor=blue,
    filecolor=blue,      
    urlcolor=blue,
    citecolor=blue,
    pdftitle={My Title},
    pdfauthor={Author Name},
    pdfsubject={Subject},
    pdfkeywords={Keywords}
}

\begin{document}
\title{Structured Spectral Compression based Low-Bitrate Secure Speech Communications for Internet of Things assisted Non-Terrestrial Networks}

\author{Li Ping Qian, \emph{Senior Member, IEEE}, Zhehan Chen, Qianru Wang, \emph{Graduate Student Member, IEEE}, Qian Wang, \emph{Member, IEEE}, 
Yuan Wu, \emph{Senior Member, IEEE}, and Xuemin Sherman Shen, \emph{Fellow, IEEE} 
}



\maketitle

\begin{abstract}
The Internet of Things assisted Non-Terrestrial Network (IoT-NTN) has played a crucial role in satellite communications, emergency communications,  and even military communications. However, the severely limited bandwidth resources of the narrow-band IoT-NTN bring a great challenge in voice services. Meanwhile, the widespread application of wideband speech data has also exacerbated the concern on the transmission delay  and privacy disclosure. To conceal the personal privacy carried in the speech data while simultaneously conserving spectrum resources and enhancing the efficiency and quality of speech communications, this paper focuses on the \underline{L}ow-\underline{B}itrate \underline{S}ecure \underline{S}peech \underline{C}ommunications based on the \underline{S}tructured \underline{S}pectral \underline{C}ompression (LB-S2C$^2$). Specifically, the Mel spectral matrix of the speech signal is first encoded at the transmitter side through compressive sensing based on waveform segmentation and data quantization. Then, the Automatic Repeat Request (ARQ) is combined with forward error correction to achieve reliable transmission of speech signals over wireless channels. Thirdly, the received signals are recovered as the speech at the receiver side. Finally, we conduct a series of simulation experiments for the performance evaluation of LB-S2C$^2$. Our simulations reveal that the dictionary matrix used for the speech reconstruction is different from the one used for the high-order matrix sparsification by even only approximately 0.1\%, and then the accurate speech recovery fails. It implies that the speech data can be securely transmitted when the dictionary matrix is preserved. More importantly, the LB-S2C$^2$ exhibits a very high privacy protection capability with the average voiceprint similarity to be only 0.3, which is much lower than the 0.8 of the semantic speech communication scheme DeepSC-S, and even lower than the 0.33 of the latest speech communication scheme OFI-OFCNB. In addition, our simulations reveal that the proposed structured speech coding boasts a time complexity of merely $O(n)$, and the proposed speech recovery scheme requires the 12-bit memory storage only, which outperforms the traditional encryption algorithms proposed for speech communications. In comparison with the conventional compression techniques, our spectral compression method renders the coding rate of only 3.9kbps, which is lower than the current lowest speech coding rate of 6.3kbps achieved by G.723. 
\end{abstract}

\begin{IEEEkeywords}
Speech communication, structured spectral compression, data security,  Mel-spectrum reconstruction.
\end{IEEEkeywords}

\section{Introduction}

With the explosive growth of Internet of Things (IoT) services, the Non-Terrestrial Network (NTN) has become appealing resolutions for seamless coverage in the next-generation wireless transmission, by integrating non-terrestrial communication technologies such as satellites, drones, and balloons \cite{ref1}. 
Especially, the IoT-NTN, which has been standardized in the 3rd Generation
Partnership Project (3GPP), offers global satellite IoT services particularly in scenarios like agricultural monitoring, environmental surveillance, and post-disaster recovery \cite{ref2,ref3}. It significantly enhances the capabilities of a wide range of IoT applications, such as global
asset tracking, for low-power and low-bandwidth equipments.   However,  many practical and intense demands exist for voice services over the IoT-NTN, and it remains  a blind spot due to a huge bandwidth gap, only requiring around 3.75 Kbps bandwidth resources according to the
3GPP requirements \cite{ref4,ref5}. Such low bandwidth of the IoT-NTN results in the  struggle with voice services, especially for high-quality real-time speech communication  in remote locations where traditional connectivity may be limited. For instance, in deep-sea scientific research and operation scenarios, traditional speech communication methods suffer from three core shortcomings: sensitivity to channel distortion (signal degradation caused by multipath effects), high resource consumption for encryption, and weak anti-eavesdropping capability. These limitations make it difficult for researchers to maintain real-time speech communication with surface teams to discuss experimental data, respond to emergencies, or adjust plans. Traditional communication approaches face irreconcilable trade-offs among channel adaptability, security redundancy, and energy efficiency constraints. Therefore, there is an urgent need for an integrated compression-and-encryption architecture that can address the challenges of reliability, secrecy, and energy sustainability in speech communication under deep-sea conditions.


The successful standardization of low-rate coders has led to their widespread application in military and satellite communications, ranging from 6kbps down to 600bps \cite{ref6}. In the military domain, the  Mixed Excitation Linear Prediction enhanced (MELPe) coding is widely employed, offering a good speech quality at 2.4kbps with robust anti-interference, though its speech quality deteriorates significantly under extreme interference conditions \cite{ref7}. And the computational complexity of the MELPe codec is relatively high, demanding a superior device. In satellite communications,  the Advanced Multi-Band Excitation (AMBE) coding provides relatively clearer speech at bandwidths as low as 2kbps \cite{ref8}. However, the speech quality with AMBE drops markedly at very low bandwidths, and its anti-interference needs to be improved in environments with pronounced multi-path effects. In addition, some of the latest research findings can be effectively applied to narrowband communication scenarios.  Zhao et al. proposed the OFI-OFCNB system\cite{ref53}. This system significantly improves coding efficiency and real-time performance by optimizing the feedback mechanism and introducing dynamic buffer management. Experiments show that it demonstrates outstanding performance advantages in resource-constrained underwater network scenarios. Note that most existing low-rate speech communication technologies suffer from the issue of poor voice quality while achieving low-bandwidth transmission, leading to a critical question of how to balance the speech quality and compression \cite{ref9}. 
Another tricky question is that in order to compress the speech at extremely low bitrate, it typically implies that the computational overhead of the encryption algorithms for security must be minimized, sacrificing the complexity and strength of the encryption, and thereby raising concerns about the communication security \cite{ref16}.

As is well-known, the key to addressing the first question in low-rate speech communication hinges on the compression coding to use. Traditional speech coding technologies have made substantial progress so far. For example, the  Code Excited Linear Prediction (CELP) technology was first developed in \cite{ref11}-by combining the codebook search and linear prediction for mid-to-low bitrate speech compression, significantly enhancing the compressed speech quality. And the Opus codec was designed in \cite{ref10}  as an efficient audio compression technology, to offer the high-quality speech and music coding suitable for real-time communications and streaming applications. Furthermore, the proliferation of neural network-based technologies has rapidly advanced the speech compression coding. For example, \cite{ref12} developed a hybrid neural network codec called LPCNet  for speech compression, which combines the linear predictive coding with deep neural networks to achieve a low-bitrate speech coding, and significantly improves the compression efficiency and sound quality. The Google team proposed the ultra-low-bitrate speech compression algorithm Lyra in \cite{ref13}, by utilizing a variational autoencoder (VAE)-based approach to extract the latent features of speech signals at the sending end, which achieves a minimum data transmission while maintaining a high-quality voice reconstruction. Last but not the least, the structured coding that organizes the data according to a specific structure or pattern can be also considered to improve the compression efficiency. It is initially used for image and video coding, which has shown excellent compression performance \cite{ref47}. It can be further applied to speech compression, for instance, \cite{ref15} proposed a spatial audio compression algorithm, based on a novel reformulation of the Higher-Order Directional Audio Coding (HO-DirAC) method.  Note that most existing compression coding can achieve the low bitrate and enhance the transmission efficiency, however, requiring high computational resources at the receiver side to reconstruct the voice or dissatisfying the low latency requirement in practice \cite{ref14}. Moreover, most achieve the extremely low-bitrate transmission without adding redundancy from a security perspective, thus implying the vulnerability to eavesdropping and attacks in communication.

To address the second question on the security issue in speech communications, most existing studies are performed to reduce the physical space on the various storage media and reduce the time of sending data over the Internet with a complete guarantee of encrypting the data and hiding it from intruders \cite{ref20}. Many well-performing data encryption algorithms have been proposed so far, to resist the eavesdropping, tampering, and illegal authentication \cite{ref17}. For example, the Blowfish algorithm as a symmetric encryption algorithm, which is widely applied for its efficiency and robust security, e.g., in \cite{ref18} for secure cloud data storage system. The Rivest-Shamir-Adleman (RSA) cryptography algorithm was combined with the compression steganography techniques in \cite{ref19}, which was used to decrease the amount of every transmitted data aiding fast transmission while using slow Internet or take a small space on different storage media. Note that most existing encryption algorithms typically introduce additional computational and storage overhead to guarantee the security. That is, the encryption-the speech data into the ciphertext, potentially increasing the total data volume and consequently increasing the bandwidth and latency required  for transmission. Additionally, the key management and the decryption can also incur extra costs, for example, consuming a certain computing cost for speech decoding \cite{ref48}. All these factors will definitely compromise the real-time transmission and voice service quality.

In addition, to address security vulnerabilities and privacy leaks in the Internet of Things (IoT), Xue et al. proposed the BareDFU system\cite{ref49,ref52}, which systematically revealed security issues in Appified IoT devices throughout the process, including authentication bypass and firmware tampering. They successfully identified and validated six types of DFU vulnerabilities in commercial devices. Han et al. proposed a system named QKNet\cite{ref50,ref51}. By integrating the Hardware Security Module (HSM), tamper-proof technology and quantum Key Distribution (QKD) technology, an end-to-end security framework covering the physical layer, network layer and application layer was constructed, achieving full life cycle protection for LPWA Internet of Things devices. However, the aforementioned schemes also incur a large amount of additional computational overhead, which can impact the real-time communication performance in narrowband environments.

Nowadays, novel speech semantic  communication systems have been intensely researched by integrating the advantages of classical transmission scheme and deep learning,  to selectively extract, compress and transmit  the semantic level information to communicate \cite{ref21,ref22,ref23}. Specifically, Qin et al.  designed a speech-to-speech semantic communication system called DeepSC-S in the field of audio signal transmission, which  improved the accuracy of voice recovery by assigning higher weights to the important features during feature extraction \cite{ref21}. Based on this,  DeepSC-SR was further developed by extracting the text-related semantic features at the sending end and restoring the text transcriptions at the receiving end, which can significantly reduce the data amount for transmission and maintain a good performance in speech recognition \cite{ref22}. Niu et al. proposed a new coding method which maps the source of speech signals onto a semantic latent space, thereby increasing the coding gain and saving up to 75\% of channel bandwidth costs \cite{ref23}. Overall, these semantic methods significantly enhance the transmission efficiency while guaranteeing the speech recovery quality. But most of them rely on the servers with powerful computing capability and often overlook the communication security issues, which are thus not applicable to low-power and low-bandwidth equipments in emergency or military communications. Table \ref{table1} shows the advantages and limitations of the above-mentioned low-rate encoders, encryption algorithms and semantic communication methods.

\begin{table}[!t]

  \caption{Advantages and Limitations of Different Methods}
  \label{table1}
  \centering
  \begin{tabular}{|c||c||c|}
  \hline
  Method & Advantages & Limitations\\
  \hline
  MELPe & Low rate & No privacy protection\\ 
  \hline
  AMBE & Ultra-low rate & No privacy protection\\
  \hline
  Blowfish & Real-time encryption & Large bandwidth\\
  \hline
  RSA & Asymmetric encryption & High computational cost\\
  \hline
  DeepSC-S & Ultra-low rate & Privacy leakage\\
  \hline
  \end{tabular}
\end{table}

Taking into account the balance between low bitrate and high security as discussed above, this paper thus proposes a low-bitrate and secure speech communication system based on the  structured spectral compression coding, referred to as $ \text{LB-S2C}^{2} $ for short. Note that this kind of structured spectral compression coding can be also used for secret imaging sharing \cite{ref24}.
Specifically in our $ \text{LB-S2C}^{2} $, at the transmitting end, the mel spectrogram matrix of the speech signal is encoded into a sparse matrix through the waveform segmentation and high-order matrix sparsification based on the compressive sensing. At the receiving end, the decompression and post-processing are performed  on the received compressed data to reproduce the high-fidelity audio.  The main contributions of this paper are summarized as follows:

\begin{itemize}
	\item At the transmitter side of $ \text{LB-S2C}^{2} $, the spectrum matrix is first extracted and then compressed, by the high-order sparse structured compression in combination with the waveform segmentation. And the compressed spectral data undergoes the quantization to further reduce its memory footprint,  which can dramatically reduce the required transmission bandwidth. This thus results in a coding rate of merely 3.9kbps, which is only equivalent to 6.0\% of G.711 and 48.8\% of G.729. Our $ \text{LB-S2C}^{2} $ with this structured spectral compression coding can be regarded as a specialized symmetric encryption system, with a time complexity of merely $O(n)$ and a memory requirement of just 12 bits. Unlike other encryption algorithms, it incurs no additional computational or storage overhead, thereby conserving computational and storage resources. 
	\item At the receiver side of $ \text{LB-S2C}^{2} $,  the speech is restored based on the mel-spectrum reconstruction through separating the mel-spectrum dimensions and sparse vectors. To improve the restored speech quality, the noise is reduced based on the singular spectrum analysis. Furthermore, the dictionary matrix involved for recovery exhibits a high sensitivity to variations, that is, even a 0.1\% change in data can decrease the peak signal-to-noise ratio (PSNR) of the reconstructed spectrum by more than 20dB. This implies that our system can effectively prevent tampering with the transmitted data, to ensure the strong voice security.
More importantly, it exhibits a very high privacy protection capability with the average voiceprint similarity to be only 0.30, which is much lower than that of most popular deep learning-based semantic speech communication schemes.
	\item We build up an experimental demonstration to verify the performance of $ \text{LB-S2C}^{2} $. The experimental results indicate that the speech recovered by the $ \text{LB-S2C}^{2} $ yields a mean opinion score (MOS) value about 4.12. It implies that our $ \text{LB-S2C}^{2} $ is entirely capable of extracting useful and important information in speech, and achieving a reliable data transmission and a high-fidelity speech recovery. Furthermore, our simulations reveal that since it is intractable to decipher the dictionary matrix used for the speech reconstruction, the speech data can be securely transmitted when the dictionary matrix is preserved. More importantly, the $ \text{LB-S2C}^{2} $ exhibits a very high privacy protection capability with the average voiceprint similarity to be only 0.3, which is much lower than that of most popular deep learning-based semantic speech communication schemes. Also, in comparison with the conventional compression techniques, our spectral compression method renders the coding rate of only 3.9kbps, which is lower than the current lowest speech coding rate of 6.3kbps achieved by G.723. 

\end{itemize}

The rest of the paper is organized as follows: Section II presents the framework of proposed $ \text{LB-S2C}^{2} $ system and design details. The security of speech communication is analyzed in Section III. Section IV primarily shows the experimental demo and performance evaluation of $ \text{LB-S2C}^{2} $ system.. Section V concludes the paper.

\section{$ \text{LB-S2C}^{2} $ system design}
In this section, we first present the framework of the low-bitrate secure speech communication system based on the structured spectral compression, referred to as $ \text{LB-S2C}^{2} $. Then, we present the details of designing the $ \text{LB-S2C}^{2} $ system.

  \vspace{-0.12in}
\subsection{Framework of $ \text{LB-S2C}^{2} $ System}
\begin{figure}[h!]
  \centering
  \includegraphics[width=\linewidth]{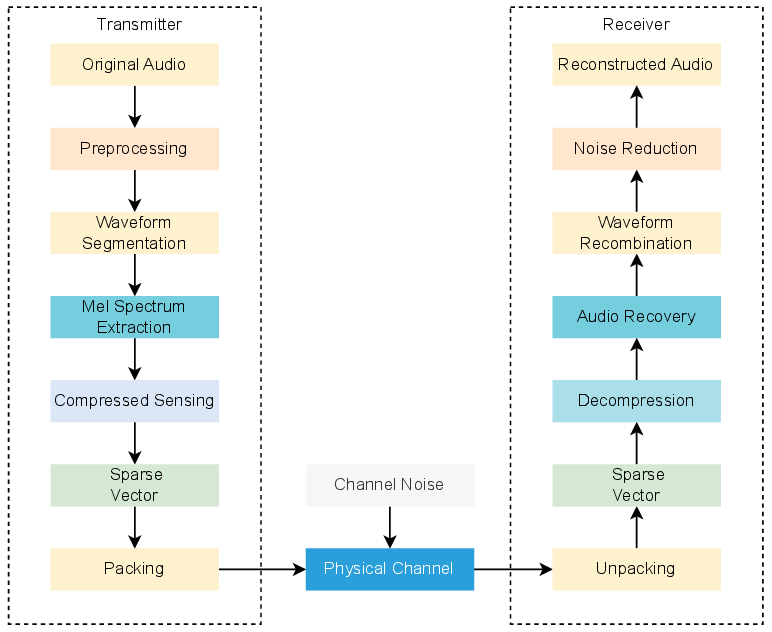}
  \vspace{-0.1in}
  \caption{The overall frame diagram of $ \text{LB-S2C}^{2} $ system.}
  \label{fig:fig1}
  \end{figure}

We present the $ \text{LB-S2C}^{2} $ system to realize the low-bitrate secure speech communication based on the structured spectral compression. The overall framework of $ \text{LB-S2C}^{2} $ system is depicted in Fig. \ref{fig:fig1}. At the transmitter side, the waveform segmentation and the mel spectrum extraction followed by the structured encoding are combined to structurally compress the spectrum of the original audio for transmission. At the receiver side,  the spectrum reconstruction and the audio recovery followed by the noise reduction are employed to reconstruct the speech. Specifically, the system input  is a time-domain sequence $\mathbf{s} = [s_1, s_2, \ldots, s_N]$ obtained by sampling the speech signal, where $N$ represents the number of sample points. This sequence is first subjected to preprocessing (including volume normalization and silent segment trimming) and waveform segmentation at the transmitter side. Subsequently, the segmented sequences undergo the mel-spectrum extraction, and then the extracted mel spectrogram is fed into a structured encoding module for sparsification. After that, encoded data is then packeted and transmitted over the physical channel to the receiver. Finally, the received data is recovered as the denoised  time-domain waveform $\mathbf s^{\prime}$ corresponding to the reconstructed speech through the spectrum reconstruction, audio restoration, and waveform reshaping. The entire process  can be denoted as the following equations:
\begin{equation}
\mathbf s^{\prime}=\mathrm{\mathbb{D}(\mathbb{R}_c(\mathbb{R}(\mathbb{F}_d(\mathbb{C}(\mathbb{F}_c(\mathbb{M}el(\mathbb{S}(\mathbb{P}(s)))))))))},
\end{equation}
where $\mathbb{D}(\cdot)$, $\mathbb{R}_{\mathrm{c}}(\cdot)$, $\mathbb{R}(\cdot)$, and $\mathbb{F}_{\mathrm{d}}(\cdot)$ denote the denoising module, waveform reassembly module, audio recovery module, and spectrum reconstruction module at the receiver, respectively, while $\mathbb{C}(\cdot)$ represents the physical channel. Here, $\mathbb{F}_{\mathrm{c}}(\cdot)$, $\mathbb{M}el(\cdot)$, $\mathbb{S}(\cdot)$, and $\mathbb{P}(\cdot)$ mean the structured encoding module, spectrum extraction module, waveform segmentation module, and preprocessing module, respectively, at the transmitter side. It is worth noting that we do not assume any specific distribution to the fading and noise of physical channel.


  \vspace{-0.12in}
\subsection {Design of $ \text{LB-S2C}^{2} $ System}

\begin{figure*}[h!]
  \centering
  \includegraphics[width=\linewidth]{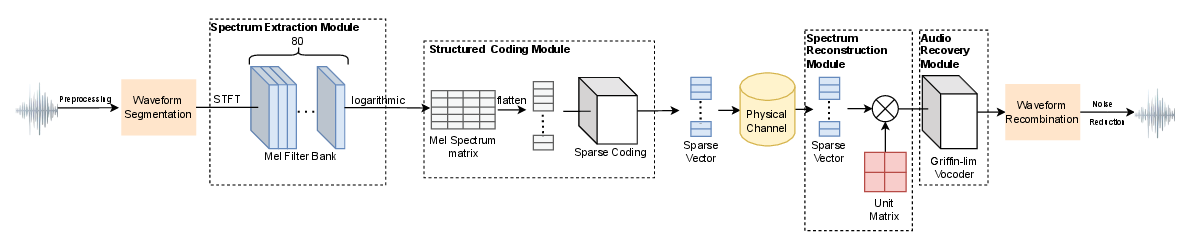}
  \vspace{-0.15in}
  \caption{The detailed diagram of  $ \text{LB-S2C}^{2} $ system architecture.}
  \label{fig3}
  \end{figure*}

The detailed diagram of the whole system architecture is shown in Fig. \ref{fig3}, where the structured encoding method and speech reconstruction in $ \text{LB-S2C}^{2} $ are based on the compressed sensing theory. Specifically, the  $ \text{LB-S2C}^{2} $ system consists of the pre-processing, mel-spectrum extraction and structured compression modules at the transmitter side, and the spectrum reconstruction, audio recovery and noise reduction modules at the receiver. In the following, we interpret the implementation and functionality of each module in details.

\textit{1)Structural Spectrum Compression based Speech Coding}

Note that the audio preprocessing and structural encoding are employed to propose the structural spectrum compression-based speech coding at the transmitter side. And the compressive encoding performance directly impacts the bitrate of data transmission. We will provide a detailed description of the function of each module at the transmitter side.

$\bullet$ \textit{Preprocessing}

Specifically, the input audio is first sampled  with the sampling rate  set to be 16000, and then followed by  volume normalization and removal of extended silent segments. For the volume normalization, we first  calculate the root mean square of the audio samples and  convert it to decibels. The difference of this from the target decibel level is then used to decide whether the volume should be increased or decreased. Finally, the volume is adjusted to the desired value given by 
\begin{equation}
\mathbf{\hat{s}}=10^{\frac{s_{1}-20\log_{10}\sqrt{\frac{\Sigma_{1}^{N}s^{2}}{N}}}{20}}\mathbf{s},
\end{equation}
where $N$ is the length of the time series, $s_{1}$ denotes the target decibel level of the waveform amplitude, and $\mathbf{\hat{s}}$ denotes the waveform sequence after volume normalization.
Note that this volume normalization not only standardizes the volume levels but also prevents  distortion caused by inconsistent audio intensities.

For the removal of  extended silent segments,  we first crop the audio to make its length an integer multiple of the detection window size, and convert the floating-point waveform to a 16-bit mono Pulse Code Modulation(PCM) format. Then, the speech detection is performed, marking out active and inactive speech regions. Subsequently, we utilize a moving average to smooth the speech detection results, and expand the speech regions to ensure that the edge portions are not inadvertently removed. Ultimately, this process yields the audio waveform data with prolonged silent segments excised.

$\bullet$ \textit{Waveform Segmentation}

To enhance the spectral compression efficiency, we employ the waveform segmentation to process the time-domain waveform of the preprocessed audio. This significantly reduces the size of the sensing matrix required for a single processing step in orthogonal matching pursuit algorithms, that is compressed to $\frac1{400}$ of its original size, thereby significantly shortening the compression time and effectively boosting the compression performance. 

 Specifically, the audio waveform is divided into $n$ contiguous and equally-sized sub-waveforms, where we set $n=20$ here.  The segmentation is expressed as
\begin{equation}
\mathbf{\hat{s}}(t)=\sum_{i=1}^n\int_{(i-1)\Delta t}^{i\Delta t}\mathbf{\bar{s}}(t')\chi_{[i-1,i]}(t')dt',
\end{equation}
where $\mathbf{\hat{s}}(t)$ denotes the normalized time-domain waveform, and $\mathbf{\bar{s}}(t^{\prime})$ represents the sub-waveform obtained after segmentation. The indicator function $\chi_{[i-1,i]}(t^{\prime})$ involved has a value of 1 within the time interval $[(i-1)\Delta t, i\Delta t]$ corresponding to the $i$-th sub-waveform, and 0 otherwise, thus precisely delineating the range of each sub-waveform. And $\Delta t$ denotes the duration of each sub-waveform, ensuring the uniformity in the waveform segmentation. These segments are then sequentially fed into the mel-spectrum extraction module for further processing.

$\bullet$ \textit{Mel Spectrum Extraction}

After waveform segmentation, the mel spectrum is extracted for each sub-waveform $\mathbf{\bar{s}}$ in turn. Extracting the mel spectrum primarily involves two steps. The first step is to compute the amplitude spectrum of the audio using the short-time Fourier transform (STFT) \cite{ref33}. The STFT process entails taking a finite-time window function $h(t)$ to multiply with the speech signal, and incrementally computes the time-related spectrum as the window slides with time in hop size. The STFT of the sub-waveform $\mathbf{\bar{s}}$ can be defined as
\begin{equation}
\mathrm{STFT}(\tau,f)=\int_{-\infty}^{\infty}{\mathbf{\bar{s}}(t)h(t-\tau)e^{-j2\pi ft}dt},
\end{equation}
where  $\mathrm{STFT}(\tau,f)$ denotes the spectrum result when the window is centred at $\tau$.
The specific process involves windowing and framing the speech signal, and stacking the STFT results of each frame along another dimension to generate a linear spectrum.

The second step is to map the amplitude spectrum using a mel filterbank to ultimately obtain the mel spectrum \cite{ref34}.  Since the resulting linear spectrogram is relatively large, it is typically transformed through a mel-scale filterbank to obtain a mel spectrum of appropriate size for audio feature representation. Since the human hearing is more sensitive to low Hz and less sensitive to high Hz,  converting Hz frequencies to mel frequencies is imperative, which can render the perception of frequency to be linear. Specifically, the transformation  is formulated as
\begin{equation}
\mathbf{L}=2595\log_{10}(1+\frac{{\mathrm{STFT}}}{700}),
\end{equation}   
where $\mathbf{L}$ represents the mel spectrum matrix after conversion. To simulate the human ear's suppression of high-frequency signals, a mel filterbank consisting of a set of usually 80 triangular filters is applied to the linear spectrum to extract low-dimensional features. This step thus  emphasizes the low-frequency part while attenuating the high-frequency part, ultimately yielding the mel spectrum matrix.

$\bullet$ \textit{Sparse Coding}

After obtaining the mel spectrum of the audio, it is input into the structured encoding module for spectrum compression. We use the sparse coding as an iterative algorithm for sparse representation and feature selection  \cite{ref25,ref26,ref27}.\ By gradually selecting the most relevant basis vectors (i.e., atoms), we can achieve a sparse representation of the signal and accomplish the compression encoding.\ The algorithmic steps are given as follows.

$\textit{Residual initialization}$. This step is to initialize the residual $\mathbf{r}_0$ as a column  vector $\mathbf{f}$ flattened from the mel spectrum $\mathbf{L}$, and set the dictionary matrix $\mathbf{D}$ to the banded matrix $\mathbf{B}_d$ with each dimension size equal to the length of the flattened one-dimensional mel spectrum. In addition, we will set an artificial residual threshold $\epsilon$ to be used as a condition for iteration termination.

$\textit{Atom index determination}$. Specifically, by computing the inner product of the residual matrix obtained in each iteration with the dictionary matrix $\mathbf{D}$ atom by atom, the atom index (i.e., the column index) of the dictionary matrix is returned for the maximum inner product. And the set of indices is updated accordingly. This can be formulated as
\begin{align}
\lambda_{k}&=\arg\max_{j\not\in\Lambda_{k-1}}|<\mathbf{d}_{j},\mathbf{r}_{k-1}>|,\nonumber\\
\Lambda_k&=\Lambda_{k-1}\cup\lambda_k,
\end{align}
where $\Lambda_k$, $\mathbf{d}_{j}$,  $\mathbf{r}_{k-1}$, and $\lambda_k$ indicate the set of index values at the $k$-th iteration, the $j$-th column of the dictionary matrix $\mathbf{D}$, the the residual $\mathbf{r}$ at the $(k-1)$th iteration, and the chosen atom index at the k-th iteration, respectively. \ The operations $|<\cdot>|$ and $\cup$ mean the inner product and the union operation, respectively. This step determines the dictionary element which best explains the residual in the current iteration.

$\textit{Sparse coefficient computation}$. We determine the sparse coefficient vector by using the least squares, mathematically expressed as
\begin{equation}
\mathbf{p}_k = \arg\min_{\mathbf{x}} \|\mathbf{f} - \mathbf{D}_{\Lambda_k} \mathbf{x}\|_2,
\end{equation}
where $\mathbf{p}_k$ and $\mathbf{D}_{\Lambda_k}$ indicate the sparse vector which enables the least square at the $k$-th iteration\ (i.e., the output of this compression algorithm) and the column subset of dictionary matrix $\mathbf{D}$ selected according to the index set $\Lambda_k$, respectively.\ The operation $||\cdot||_2$ means the Euclidean 2-norm operation. This step aims to find an optimal sparse vector $\mathbf{p}_k$, such that the dictionary $\mathbf{D}$ weighted by sparse coefficients in $\mathbf{p}_k$ can most closely reconstruct the original vector $\mathbf{f}$. That is, we have
\begin{align}
\mathbf{f}_k = \mathbf{D}_{\Lambda_k}\mathbf{p}_k,
\end{align}
where $\mathbf{f}_k$ represents the approximate vector of $\mathbf{f}$.

$\textit{Residual matrix reconstruction}$. Based on eq.\ (8) obtained in the above step, the residual vector of each iteration can be obtained and formulated as 
\begin{align}
\mathbf{r}_k = \mathbf{f} - \mathbf{f}_k,
\end{align}
where $\mathbf{r}_k$ represents the residual at the $k$th iteration.

$\textit{Iterating}$. This step is to compare the L2 norm of the residual matrix with our predefined threshold $\epsilon$ to determine whether the iteration should be terminated. When the final $k$ iterations are completed, the resulting sparse vector $\mathbf{p}_k$ is the compressed mel spectrum we want.

In brief, the sparse coding algorithm with the above steps is summarized in \textbf{Algorithm 1}.  In this way, the sparse coding realizes the speech structured compression. Before transmission through the channel, we extract the non-zero values and their corresponding position information from the sparse matrix $\mathbf{V}$ formed by stacking sparse vector $\mathbf{p}$ row by row. These are then packaged with size information of the spectrum and sent to the receiver for speech reconstruction.

\begin{algorithm}
    \caption{Sparse coding}
    \label{euclid}
    \SetKwInOut{Input}{Input}
    \SetKwInOut{Output}{Output}
    \Input{$\mathbf{D},\mathbf{f},\epsilon$}
    \Output{$\mathbf{p}_k$}
    \BlankLine
    \textbf{Initialization:} $\mathbf{r}_0 = \mathbf{f}$, $\Lambda_0 = \emptyset$, $\epsilon = 0.5$, $\mathbf{D} = \mathbf{B}_d$  \\
    \For{$k = 1, 2, \dots$}{
        \textbf{Step-1:} $\lambda_k = \arg \max_{j \notin \Lambda_{k-1}} |<\mathbf{d}_j,\mathbf{r}_{k-1}>|$\tcp*{Select atom with maximum correlation}
        \textbf{Step-2:} $\Lambda_k = \Lambda_{k-1} \cup \{\lambda_k\}$\tcp*{Update index set}
        \textbf{Step-3:} $\mathbf{p}_k = \arg \min_\mathbf{x}||\mathbf{f} - \mathbf{D}_{\Lambda_k}\mathbf{x}||_2$\tcp*{Sparse coefficient estimation}
        \textbf{Step-4:} $\mathbf{f}_k = \mathbf{D}_{\Lambda_k}\mathbf{p}_k$\tcp*{Reconstruct signal}
        \textbf{Step-5:} $\mathbf{r}_k = \mathbf{f} - \mathbf{f}_k$\tcp*{Update residual}
        \textbf{Step-6:} \If{$\|\mathbf{r}_k\|_2 \leq \epsilon$}{terminate the algorithm}
    }
\end{algorithm}

\textit{2)Reliable Transmission}

The transmission scheme adopted here relies on the hybrid automatic repeat request (HARQ) mechanism \cite{ref35}. It is because that in practical implementation, HARQ as an advanced transmission protocol not only enhances the data transfer reliability through the adaptive retransmission strategies, but also effectively detects the errors to ensure the integrity and accuracy of the speech data received. Additionally,  the advanced data compression incorporated  helps reduce the data volume to be transmitted, maximizing the efficiency of HARQ-based transmission. Lastly, the HARQ-based transmission scheme enables the rapid response and processing of data transfer requests, minimizing the latency while guaranteeing the quality of service. 

\begin{figure}[h!]
  \centering
  \includegraphics[width=\linewidth]{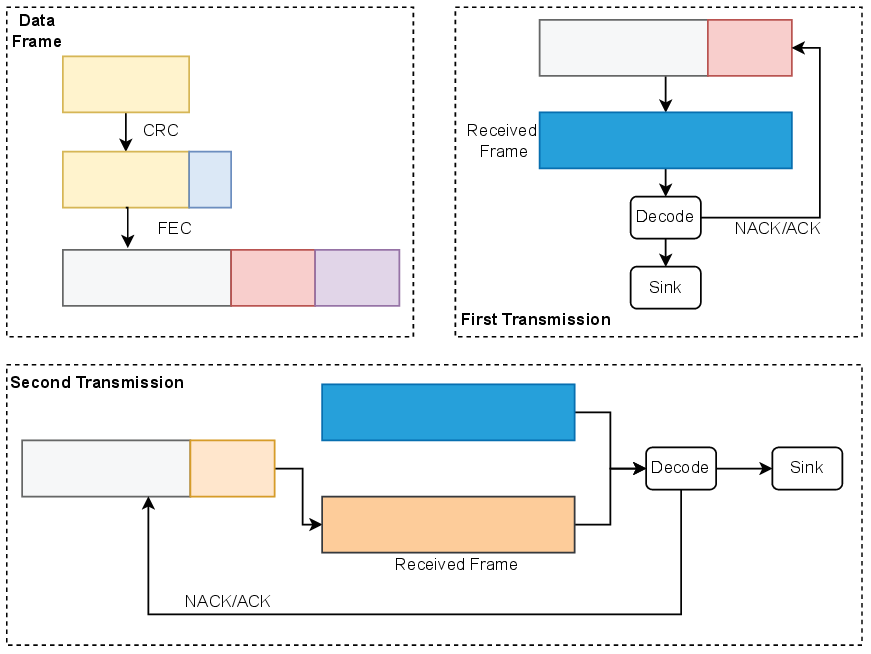}
  \vspace{-0.15in}
  \caption{The workflow of HARQ transmission in our system.}
  \label{fig4}
  \end{figure}

Fig. \ref{fig4} illustrates the workflow of HARQ transmission in our system.  Initially, the original data is uniformly divided into data packets, with each packet appended with a CRC-16 checksum to ensure the data integrity. Subsequently, these data packets undergo the Reed-Solomon (RS) encoding to enhance the resistance against burst errors. Next, the processed packets are sent one by one. Upon receiving correct data frames, the receiver promptly sends back an ACK confirmation message, allowing both the transmitter and receiver to monitor the status of data transmission in real-time. When the transmitter receives the ACK  for a particular data frame, it suspends the retransmission of that frame until all data frames have been confirmed. The receiver, meanwhile, orderly reassembles the data packets based on their numbering information within the frames, eventually restoring the original data. Note that through this  data packeting and framing strategy,  it can reduce  the link burden and significantly lower  the packet loss rate with immediate packet loss detection and retransmission. More importantly, the use of RS encoding greatly enhances the error correction capability of data transmission, thereby boosting the overall transmission efficiency.

In brief, the RS encoding-based HARQ transmission scheme used in our system not only provides high-quality data transfer services but also contributes to efficient resource utilization, demonstrating an outstanding performance in our demo experiments shown later.
 
\textit{3)Mel-spectrum Reconstruction based Speech Recovery }

The Mel-spectrum Reconstruction based Speech Recovery is implemented at the receiver side. Specifically, the received compressed mel spectrum and size
information are first reconstructed into a mel spectrum, and then the mel spectrum is converted back into the recovered audio. Moreover, the noise reduction is performed on the restored audio to improve its sound quality. We will now detail the working principles of each module at the receiver side.

$\bullet$ \textit{Information Separation}

The packed information of each segment is initially unpacked at the receiver side, extracting the the non-zero values, the position information and the mel spectrum dimensions correspondingly from the sparse matrix $\mathbf{V}$. This separation process facilitates the subsequent spectral reconstruction operations. Once the separation is complete, the sparse vectors $\mathbf{p}$ restored from the non-zero values and position information are input into the spectral reconstruction module along with mel spectrum dimensions.

$\bullet$ \textit{Spectrum Reconstruction}

The spectrum reconstruction module serves as the key component for restoring the original speech data. It reconstructs the original mel spectrum  matrix based on the sparse vectors and dimensions provided, which can be viewed as an inverse of the structured encoding.\ To reconstruct a mel spectrogram using a sparse representation, the specific steps are as follows.

$\textit{Multiplication}$. By multiplying the sparse representation with a banded matrix, we can reconstruct the approximate mel spectrogram. This can be mathematically expressed as
\begin{equation}
\hat{f}=\hat{p}B_d
\end{equation}
where $\hat{p}$ and $\hat{f}$ mean the received sparse representation vector and the recovered mel spectrum in a flattened state, respectively. 

$\textit{Reshaping}$. This step is to reshape the reconstructed mel spectrogram into the shape of the original mel spectrogram. 
This shape adjustment is critical, as it preserves the structure and characteristics of the original signal.

$\textit{Non-zero analysis}$. During the decompression, we further analyze the non-zero elements in the reconstructed mel spectrogram. By examining the positions, amplitudes, and distributions of the non-zero elements, we can evaluate the reconstruction quality and potential information loss.

Through the above steps, we can effectively restore the original mel spectrogram, which is highly consistent with the data at the transmitter side. Moreover, this spectrum reconstruction heavily relies on the authenticity of the dictionary matrix, similar to the data encryption process, where we can consider the dictionary matrix as a special type of key.

$\bullet$ \textit{Audio Restoration}

\begin{figure*}[h!]
  \centering
  \includegraphics[width=\linewidth]{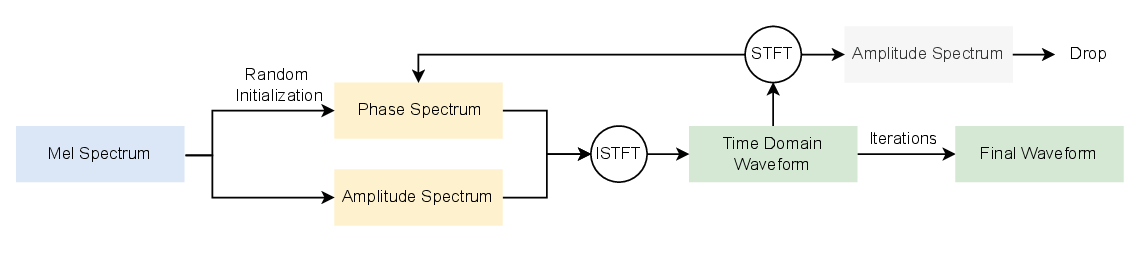}
  \vspace{-0.3in}
  \caption{The workflow for audio restoration.}
  \label{fig5}
\end{figure*}

The role of the audio restoration module is to perform a series of operations on the reconstructed mel spectrogram to convert it back into a  time-domain audio waveform. First, the linear amplitude spectrum is computed from the reconstructed mel spectrogram, by constructing a mel basis and using the non-negative least squares to find a solution, followed by applying an inverse exponential function and then raising to a power. This amplitude spectrum is then converted back into an audio waveform through the Griffin-Lim vocoder \cite{ref36}. The Griffin-Lim algorithm works as, given a known amplitude spectrum but unknown phase spectrum, iteratively generating the phase spectrum combined with the known amplitude spectrum to reconstruct the speech waveform.  The specific steps are shown in Fig. \ref{fig5}, i.e.,

    $\textit{Step1: }$ Randomly initialize a phase spectrum.

    $\textit{Step2: }$ Use the phase spectrum along with the known amplitude spectrum (derived from the reshaped mel spectrogram $\mathbf{L}^{\prime}$) to synthesize a new speech waveform via inverse STFT.

    $\textit{Step3: }$ Perform STFT on the synthesized speech to obtain a new amplitude spectrum and a new phase spectrum.

    $\textit{Step4: }$ Discard the new amplitude spectrum, and use the known amplitude spectrum combined with the new phase spectrum to synthesize a new speech.

    $\textit{Step5: }$ Repeat the above steps until the maximum number of iterations is reached, and we get the final waveform $\bar{s}_i'$.

Note that we choose to employ the Griffin-Lim vocoder rather than the  neural network-based vocoders like MelGAN or WaveGlow for audio restoration here,  due to the fact that compared to the intricate training process of neural networks, the signal processing-based Griffin-Lim vocoder  has   lower complexity and higher efficiency.

 $\bullet$ \textit{Batch Aggregation}

Since the audio restoration module outputs a series of discrete audio waveform segments, each representing a small portion of the original audio, we must aggregate these segments to reconstruct the complete audio signal. This aggregation process must be executed in the exact temporal order to ensure the coherence and temporal accuracy of the audio stream. Through such meticulous aggregation, we ultimately obtain a continuous, complete audio time-domain waveform that closely approximates the original sound quality and structure.  Specifically, this aggregation process is formulated as 
\begin{equation}
\hat{s}'(t)=\sum_{i=1}^n\bar{s}_i'(t)\chi_{[(i-1)\Delta t,i\Delta t]}(t),
\end{equation}
where $\hat{s}'(t)$ represents the reconstructed  time-domain signal, 
$\bar{s}_i'(t)$ represents the  $i$-th sub-waveform  at time $t$. 

The aggregation of audio waveform segments is an indispensable part to  achieve the desired level of fidelity. During the aggregation, we encounter the issue of boundary effects. To solve this and ensure a smooth transition, we apply the fade-in and fade-out  strategy on the edges of each sub-waveform, which effectively mitigates any abruptness that might occur at the splice points. This thus results in a smoothly concatenated audio waveform with significantly enhanced naturalness and continuity in the listening experience.

$\bullet$ \textit{Noise Reduction}

Given that the restored speech contains some background noise and exhibits glitches in the waveform that can impair its quality, we design a noise reduction module  employing the singular spectrum analysis (SSA) as its core algorithm \cite{ref37,ref38,ref39}. Through the SSA, the synthesized speech signal is first decomposed into a two-dimensional trajectory matrix. Then,  the singular value decomposition (SVD) of the trajectory matrix and selection of the most significant eigenvectors are involved  for computing the reconstruction matrix. Finally,  a denoised one-dimensional signal is reconstructed using this reconstruction matrix. The specific workflow is given as follows.

$\textit{Construction of trajectory matrix}$. The time-series data is rearranged into a trajectory matrix using a specified sliding window. The size of this matrix depends on the choice of the observation window length and delay. The trajectory matrix can be thus represented as
\begin{equation}
\mathbf{X}=\begin{bmatrix}\bar{s}'_1&\cdots&\bar{s}'_K\\\vdots&\ddots&\vdots\\\bar{s}'_l&\cdots&\bar{s}'_N\end{bmatrix},
\end{equation}
where $N$ is the length of the time series, $l$ is the window length, and $\mathbf{X}$ is an $l \times K$ matrix with $K=N-l+1$.

$\textit{SVD decomposition}$.  The SVD decomposes the trajectory matrix into eigenvectors and eigenvalues. By retaining the most important eigenvectors, the signal-to-noise separation can be achieved, that is
\begin{align}
\mathbf{X}=\sum_{i=1}^d\sqrt{\alpha_i}\mathbf{U}_i\mathbf{V}_i^T,
\end{align}
where $\alpha_i$ represents the eigenvalues of the trajectory matrix $\mathbf{X}$, $\mathbf{U}_i$ and $\mathbf{V}_i$ are orthogonal matrices, $d$ is the rank of the matrix. In this way, we obtained the decomposed eigenvectors and eigenvalues for each component of the trajectory matrix.  These decomposed components can then be used to identify and separate the underlying patterns or trends within the original time series data, enabling effective signal reconstruction and noise reduction.
]
$\textit{Grouping}$. Divide $\mathbf{X}$ into $g$ groups, and let $I_1, I_2, \dots, I_g$ be the index sets for the groups. Then the composite matrix corresponding to each group is
\begin{equation}
\mathbf{X}_{I_k}=\sum_{i\in I_k}\sqrt{\alpha_i}\mathbf{U}_i\mathbf{V}_i^\top,\quad k=1,2,\ldots,g
\end{equation}
where $\mathbf{X}_{I_k}$ is the composite matrix associated with the $k$th group.

$\textit{Signal reconstruction}$. After the grouping is completed, it is necessary to reconstruct the signals from the grouped components back into the original time series space. This step is achieved by applying diagonal averaging to the grouped matrix. For each composite matrix $\mathbf{X}_{I_k}$, the diagonal averaging operation can be expressed as
\begin{equation}
\begin{split}
y^{(i)}(n) &= \frac{1}{Q_n} \sum_{(k,j): k+j=n+1} \mathbf{X}_i(k,j),n=1,2,\ldots,N
\end{split}
\end{equation}
where $Q_n$ and $\mathbf{X}_i(k,j)$ are the number of elements on the $n$th diagonal and the element at position $(k,j)$ in the matrix $\mathbf{X}_i$, respectively.\ This operation averages the elements along the diagonals of the matrix to produce the reconstructed time series component $y^{(i)}(n)$. And the sum of all the reconstructed time series component gives the denoised speech signal, that is
\begin{equation}
s'(t)=\sum_{k\in I}y^{(k)}(n),\quad n=1,2,\ldots,N
\end{equation} 
where $I$ is the index set that contains the selected group.

Note that in practical denoising, we only need to reconstruct the sequence using the first three components with the relatively larger contributions.

\section{SECURITY ANALYSIS}

In this section, we want to analyze whether the proposed $ \text{LB-S2C}^{2} $ system can guarantee the secure speech communication. 

\begin{figure}[h!]
  \centering
  \includegraphics[width=\linewidth]{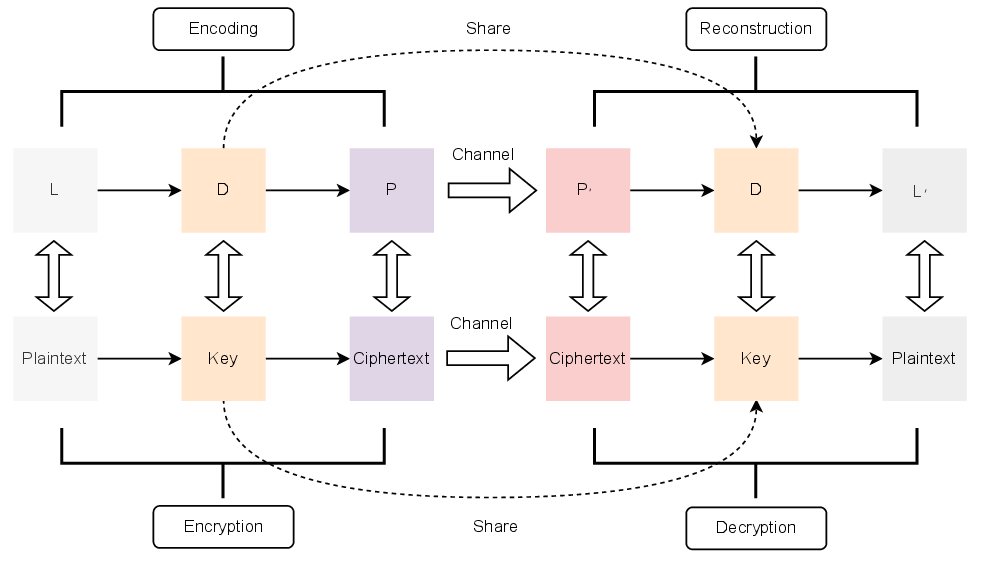}
  \vspace{-0.2in}
  \caption{Correspondence between the proposed system and an encryption algorithm system.}
  \label{fig2}
\end{figure}

Note that the speech can be recovered only when the dictionary matrix is the same as that used for the high-order matrix sparsification at the transmitter side. Therefore, we can view the proposed entire system as a special symmetric cryptographic system, as shown in Fig. \ref{fig2}. That is, the encoding process corresponds to the encryption, where the original spectrum $\mathbf{L}^{\prime}$ corresponds to the plaintext, the compressed spectrum $\mathbf{P}^{\prime}$  corresponds to the ciphertext, and the dictionary matrix $\mathbf{D}$  corresponds to the shared key. Similarly, the spectrum reconstruction process corresponds to decryption, with $\mathbf{L}^{\prime}$ and $\mathbf{P}^{\prime}$ considered after the channel transmission. Furthermore, considering the randomness of dictionary matrix, it is intractable to decipher except for by the brute force. It implies that the speech data can be securely transmitted when the dictionary matrix is preserved.

\section{SIMULATION RESULTS}

In this section, we want to evlauate the performance of proposed $ \text{LB-S2C}^{2} $ system based on the demo system. 

\subsection{Experimental Setup}

\begin{figure}[h!]
  \centering
  \includegraphics[width=\linewidth]{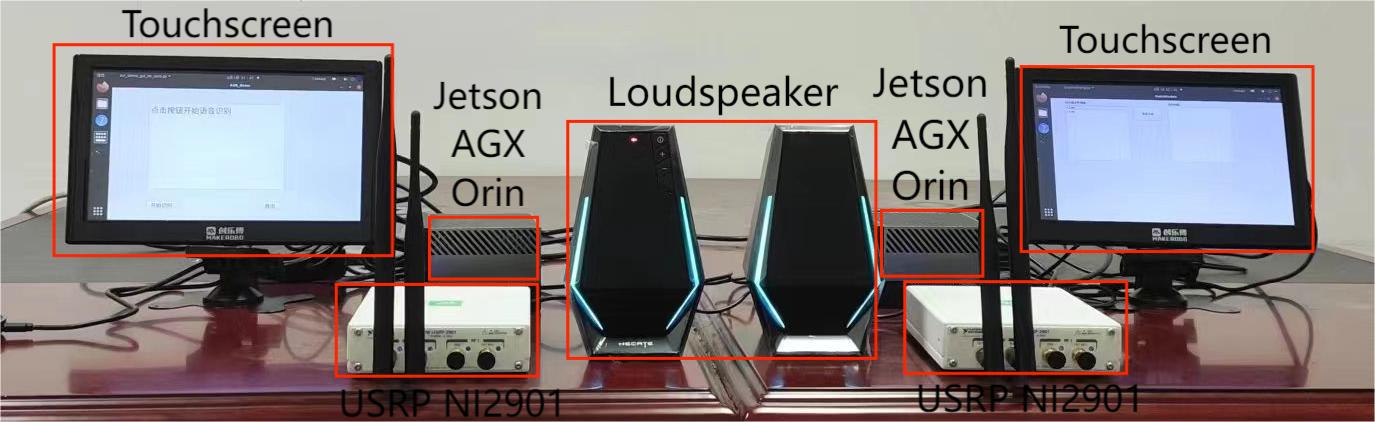}
  \vspace{-0.15in}
  \caption{Experimental system setup for speech communication.}
  \label{fig6}
\end{figure}

The system configuration for our experiments is depicted in Fig. \ref{fig6}, comprising two Nvidia Jetson AGX Orin devices, two USRP NI2901 units, two touch screens, and a speaker. The Nvidia Jetson AGX Orin serves as the primary control and computation device, while the USRP NI2901 (with a maximum output power of 20 dBm) enables the  wireless data transmission and reception. 

The operating environment for this demo is as follows: Ubuntu 20.04 LTS with real-time kernel patches is the operating system, NVIDIA CUDA 12.2 is used for gpu accelerated signal processing,the UHD 4.7.0 driver is used for USRP control,and the GNU Radio 3.9 is used for designing and debugging the wireless communication system. The data carrier center frequency is set to 830 MHz, the ACK carrier center frequency is set to 960 MHz, the USRP gain is set to 40 dB, the modulation scheme is GMSK, and the retransmission protocol is HARQ.
  \vspace{-0.1in}

  \vspace{-0.2in}

\subsection{System Performance Metrics}

The proposed system  aims to ensure the quality of reconstructed speech while enhancing the security of speech transmission, and to a certain extent, to reduce the required transmission bandwidth and increase the spectral utilization. Therefore, the performance metrics considered here primarily serve to evaluate the transmission efficiency and security, the audio compression and   restoration capabilities of the communication system.

\textit{1) Information Entropy}

From a security perspective, information entropy plays a critical role in ensuring the robustness and reliability of cryptographic systems \cite{ref44}. \ The entropy refers to the measurement of randomness or unpredictability in this speech communication system.  High entropy means that key generation has strong randomness, making it difficult for attackers to predict. This unpredictability is crucial, since it makes the brute-force attacks to be significantly more difficult.\ Mathematically, the entropy of our compressed mel spectrum matrix is given by 
\begin{equation}
H(\mathbf{P}) = -\sum_{i} p(x_i) \log_2 p(x_i),
\end{equation}
where $p(x_i)$ is the probability of occurrence of each of the different elements $x_i$ in the matrix $\mathbf{P}$.\ In security applications, a higher entropy indicates a stronger resistance to the cryptanalysis while a lower likelihood of successful attacks based on prediction.

\textit{2) Zero-frequency Analysis}

Zero-frequency analysis is an analytical technique in information theory and cryptography that reveals potential weaknesses in a cryptographic system or compression algorithm, by detecting features not present in the data (i.e., zero-frequency features) \cite{ref45}. This analytical metric helps identify and address the  security vulnerabilities in system design, thereby improving the overall data security. A lower zero-frequency count typically indicates higher security in the cryptographic system or compression algorithm, as it is difficult for attackers to infer potential patterns in the system by analyzing features that do not appear. The zero-frequency metric can be calculated as
\begin{equation}
F_z=\sum_{i=a}^b\mathbb{I}(f(i)=0),
\end{equation}
where $\mathbb{I}$ represents an indicator function with a value of 1 if the condition is true and 0 otherwise, $f(i)$ represents the frequency of occurrence of the value $i$ in the compressed mel spectrum matrix $\mathbf{P}$, $a$ and $b$ denote the minimum and maximum range of value $i$ to be analyzed(in the compressed mel spectrum matrix $P$), and $F_d$ represents the total number of values whose occurrence frequency is zero in the range.

\textit{3) Voiceprint Similarity}

As an efficient means of identity verification, voiceprint recognition plays a crucial role in the field of privacy protection \cite{ref46}. We thus proceed to rigorously assess the system capability in upholding speaker privacy, focusing specifically on the speaker verification. To this end, we choose the voiceprint similarity as the main evaluation metric, which is derived by computing the cosine similarity between the pairs of voiceprint signals. This measure spans a range from -1 to 1, where a higher value denotes an increased similarity between voiceprints. Furthermore, when this value exceeds a predefined threshold e.g., set as 0.5 in our experiment, it is inferred that the two voiceprints originate from the same speaker. The voiceprint similarity can be calculated as
\begin{equation}
   \mathrm{Similarity} = \frac{\mathbf{a}\cdot\mathbf{b}}{\|\mathbf{a}\|\|\mathbf{b}\|},
\end{equation}
where $Similarity$ represents the voiceprint similarity, $\mathbf{a}$ and $\mathbf{b}$ represent the feature vectors extracted from the audio files of the input time-domain waveform $s$ and the reconstructed time-domain waveform $s^{\prime}$, respectively, $\cdot$ represents the dot product operation, and $\|\cdot\|$ represents the magnitude (norm) of the vectors.

\textit{4) PSNR Analysis}

In speech communications, PSNR can be used to evaluate the extent of quality degradation between the original mel spectrum and the reconstructed one after undergoing the processes such as compression and encoding \cite{ref31}. A higher PSNR value indicates that the reconstructed mel spectrum has higher quality and less distortion. The PSNR of mel spectrum can be  calculated as
\begin{equation}
	\mathrm{PSNR(dB)}=20\log_{10}\left(\frac{S_{max}}{\sqrt{\mathrm{MSE}}}\right),
\end{equation}
where $S_{max}$ represents the maximum value in the mel spectrum, and $\text{MSE}$ representing the mean squared error between the original mel spectrum and the reconstructed one is defined as 
\begin{equation}
    \mathrm{MSE}=\frac1{mn}\sum_{i=1}^m\sum_{j=1}^n\left(\mathbf{L}(i,j)-\mathbf{L^{\prime}}(i,j)\right)^2
\end{equation}
where \(m\) and \(n\) represent the dimensions of the mel spectrum, \(\mathbf{L}(i,j)\) represents the value in the original mel spectrum, and \(\mathbf{L^{\prime}}(i,j)\) represents the value in the reconstructed mel spectrum. This formula quantifies the quality of the mel spectrum after processing, helping to assess the level of distortion introduced during the process.

\textit{5) MOS Analysis}

MOS is a standard method for assessing the speech quality, originating from the telecommunications industry's evaluation of telephone speech service quality \cite{ref30}. It provides a quantitative measure to rate the speech reconstruction quality here, with typical values ranging from 1 to 5. The detailed denotions are given as
\begin{itemize}
	\item 5: Excellent – The speech sounds very clear and natural, nearly indistinguishable from face-to-face conversation.
	\item 4: Good – The speech quality is satisfactory, although there may be slight noise or distortion present.
	\item 3: Fair – The  reconstructed speech contains noticeable noise and distortion, yet remains comprehensible.
	\item 2: Poor – The speech quality is significantly degraded, making it difficult to understand.
	\item 1: Very Poor – The speech is either barely or completely unintelligible.
\end{itemize}

In an actual MOS test, a group of listeners is asked to listen to a series of speech samples and provide their subjective evaluations of each sample's quality. The average of all scores constitutes the MOS score for that particular speech sample in our system. 

  \vspace{-0.12in}

\subsection{Introduction to the Comparison Scheme}

To comprehensively evaluate the performance of the proposed LB-S2C² system, we conduct comparative experiments against four schemes: DeepSC-S\cite{ref21}, DeepSC-SR\cite{ref22}, OFI-OFCNB\cite{ref53}, and the traditional Orthogonal Matching Pursuit (OMP)\cite{ref42}. We next briefly introduces their core architectures and technical characteristics, as shown in Table \ref{table5}.

\textit{1) DeepSC-S: Speech-to-Speech Semantic Communication}

DeepSC-S is an end-to-end semantic communication system that directly transmits speech features without the text conversion. It employs a deep neural network (DNN) to extract high-level semantic features from the raw audio, along wih assigning adaptive weights to critical components during the feature extraction. The receiver reconstructs the  speech using a generative adversarial network (GAN). While efficient for the semantic preservation, it lacks explicit security mechanisms and requires significant computational resources.

\textit{2) DeepSC-SR: Speech-to-Text Semantic Communication}

DeepSC-SR converts the speech into text-related semantic features at the transmitter and reconstructs text transcriptions at the receiver. By discarding the non-linguistic information (e.g., speaker identity), it achieves ultra-low bitrates but sacrifices voice characteristics and speaker privacy. Its architecture relies on the transformer-based encoders and decoders, which makes it unsuitable for high-fidelity voice generation scenarios.

\textit{3) OFI-OFCNB: Optimized Feedback Fountain Coding}

OFI-OFCNB enhances the speech communication reliability in resource-constrained networks (e.g., underwater acoustic channels) through the dynamic buffer management and feedback-controlled fountain coding.    It eliminates the build-up phase of traditional fountain codes, thus reducing the transmission latency.    However, it focuses primarily on transmission robustness without integrated encryption, leaving speech data vulnerable to eavesdropping.

\textit{4) OMP: Orthogonal Matching Pursuit}

OMP is a classic compressive sensing algorithm that reconstructs signals via iterative sparse approximation. We compare it with $ \text{LB-S2C}^{2} $'s structured coding module to highlight efficiency gains. Traditional OMP operates on full-spectrum data without segmentation or quantization, resulting in high computational complexity  $O(n^2)$ and limited security.

\begin{table}[!t]
  \caption{Core Architectures of Comparison Schemes}
  \label{table5}
  \centering
  \begin{tabular}{|c||c|}
  \hline
  Scheme & Core Architecture \\
  \hline
  DeepSC-S & DNN-based semantic encoder/decoder + GAN vocoder \\ 
  \hline
  DeepSC-SR & Transformer encoder + Text decoder \\
  \hline
  OFI-OFCNB & Feedback-driven fountain coding + RS codes \\
  \hline
  OMP & Iterative sparse approximation \\
  \hline
  \end{tabular}
\end{table}

\subsection{Experimental Results Analysis}

In the following, we present the experimental results analysis.  We conduct performance analysis and evaluation of the system, including: compression performance evaluation, transmission performance evaluation, noise reduction performance evaluation, and security assessment of the system. Through these comprehensive evaluations, we demonstrate that our proposed scheme successfully achieves a balance between low-bitrate efficiency and high-security requirements.

\textit{1) Compression Performance Evaluation} 

\begin{figure}[h!]
  \centering
  \includegraphics[width=\linewidth]{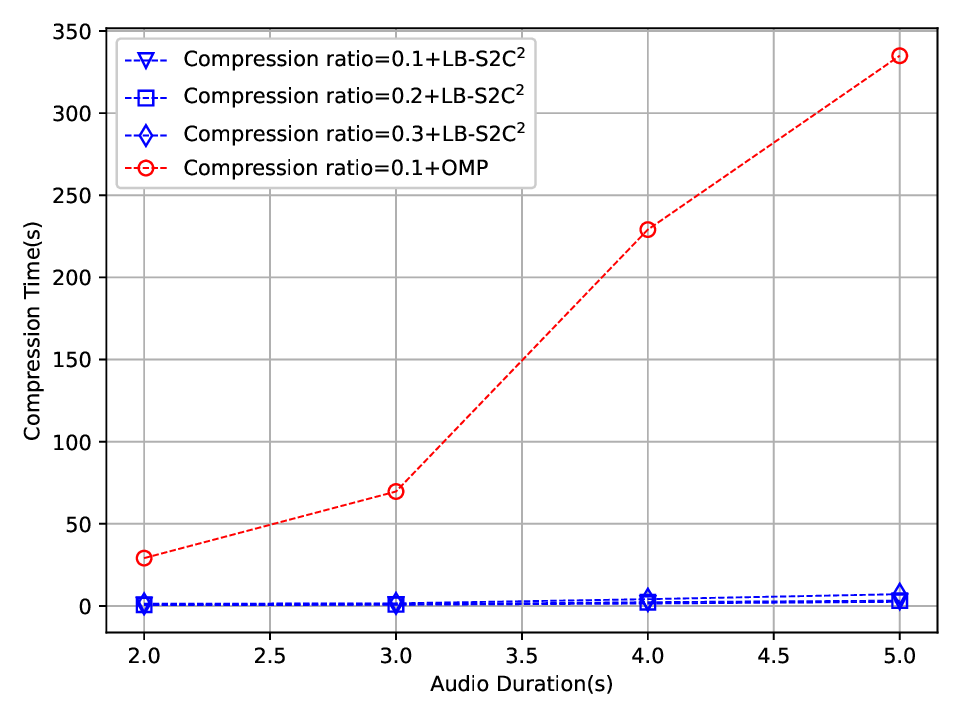}
  \vspace{-0.15in}
  \caption {Compression time comparison with conventional OMP for different audio durations.}
  \label{fig11}
\end{figure}

To evaluate the compression performance of the proposed structured encoding, we consider the performance metrics including compression time and compression ratio. Fig. \ref{fig11} shows the compression time comparison  with the traditional compressed sensing method Orthogonal Matching Pursuit (OMP) \cite{ref42}. It is shown that the  average time required to compress the Mel spectrum corresponding to the audio of different durations is reduced to be less than 10\% of the time required for OMP, indicating that our $ \text{LB-S2C}^{2} $ has a faster compression speed. Moreover, the increased compression ratio or the increased audio duration does not increase the required compression time of $ \text{LB-S2C}^{2} $, which indicates a flatly stable compression performance.

\begin{figure}[h!]
  \centering
  \includegraphics[width=\linewidth]{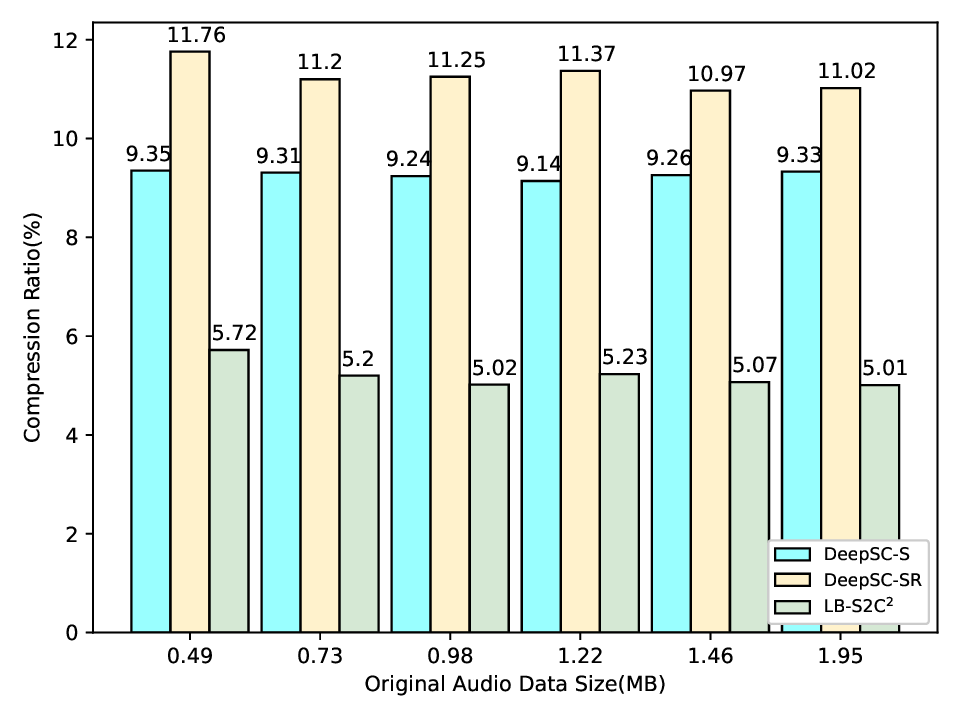}
  \vspace{-0.3in}
  \caption{Compression ratio comparison for different audio data sizes.}
  \label{fig12}
  \end{figure}

Fig. \ref{fig12} shows the compression ratio comparison with different transmitted data volumes considered. Both the DeepSC-S and DeepSC-SR semantic speech communication systems are included for benchmarking. Note that our algorithm based on the sparsification of  the mel spectrogram matrix achieves a compression ratio of only around 5\%, which is approximately 61\% of DeepSC-S and about 49\% of DeepSC-SR. This indicates that our algorithm with less data conserved for transmission can be well applied in low-power and low-bandwidth equipments in practice. 

\textit{2) Required Bandwidth Analysis}

Note that our scheme is to reduce the required bandwidth for transmission while maintaining the transmission security. In comparison with some conventional high-strength encryption algorithms (e.g., Advanced Encryption Standard with a 256-bit key size(AES-256)), our scheme avoids the increase in the data size to guarantee the security, thereby reducing the consumption of bandwidth resources. Note that the method for calculating our encoding rate is that at the sender side, we first select a certain number of 2-second, 5-second, and 10-second audio clips for compression encoding. We then calculate the ratio of the compressed data to the corresponding audio length, and take the average of these ratios to obtain our final encoding rate.

\begin{figure}[h!]
  \centering
  \includegraphics[width=\linewidth]{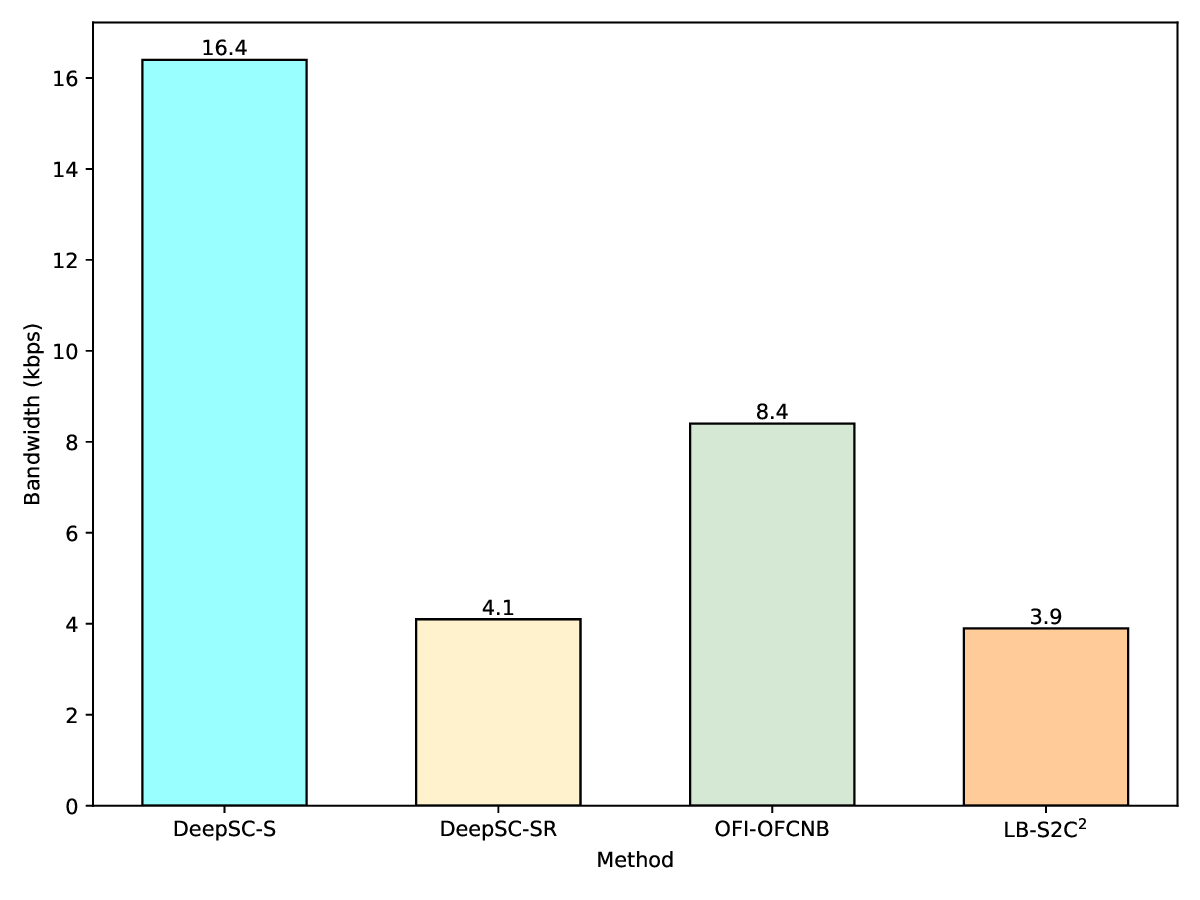}
  \vspace{-0.22in}
  \caption{Comparison of speech bandwidth required by different encoding methods.} 
  \label{fig13}
\end{figure} 

Here, we compare the speech bandwidth required by our $ \text{LB-S2C}^{2} $ with that of  DeepSC-S, DeepSC-SR and OFI-OFCNB. As shown in Fig. \ref{fig13}, our method can achieve a coding rate of only 3.9kbps, that is only about 23.8\% of the bandwidth of DeepSC-S, 95.1\% of that of DeepSC-SR and 46.4\% of that of OFI-OFCNB. This is because our method quantizes the transmitted data by integers instead of floating-point operations, under the premise of minimal impact on the speech quality. This approach effectively reduces the required  transmission bandwidth while preserving the security of the speech. It is worth noting that during data transmission, additional overhead from network protocols (RTP, UDP, and IP) will be introduced, so the actual total coding rate is 7.1 kbps.

\textit{3) Transmission Delay Analysis}

Here, we test the transmission delay of our scheme under different transmit power and distance conditions, in comparison with DeepSC-SR. For this purpose,   we selected a 6-second speech segment, encoded it using the DeepSC-SR system to generate a data matrix, and then transmit it in comparison with transmitting our spectrogram sparsification matrix, while recording the corresponding delays.

  \vspace{-0.1in}
\begin{figure}[h!]
  \centering
  \includegraphics[width=\linewidth]{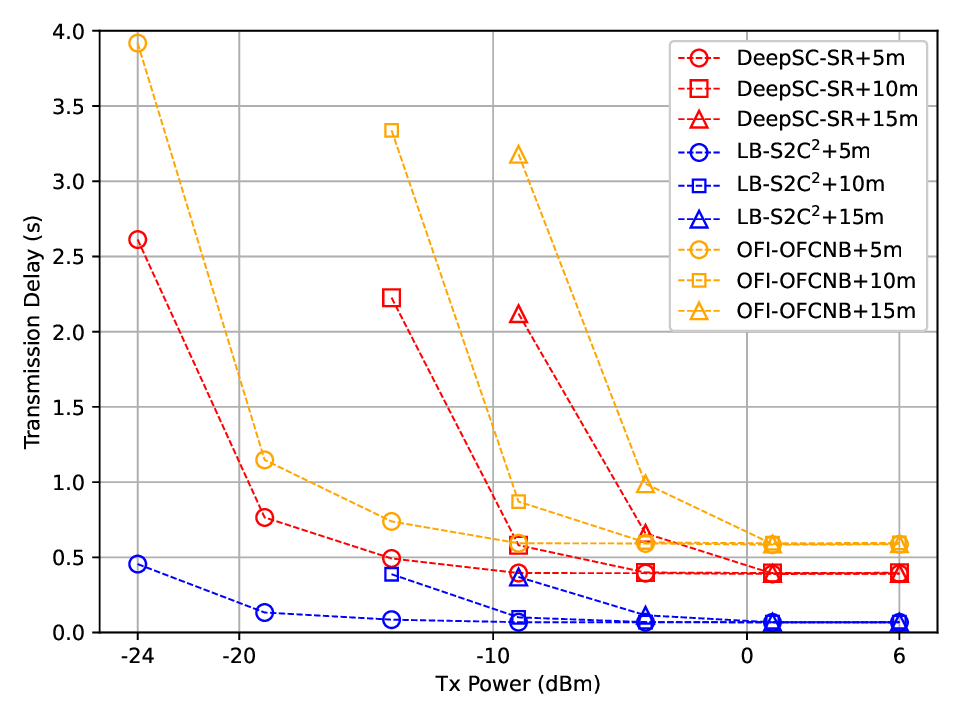}
  \vspace{-0.3in}
  \caption{Comparison of transmission delay under different conditions.}
  \label{fig14}
\end{figure}
  \vspace{-0.1in}

As shown in Fig. \ref{fig14} with the varying transmit powers and transmission distances considered, our $ \text{LB-S2C}^{2} $ achieves an average delay reduction to below 20\% of that in DeepSC-SR and 15\% of that in OFI-OFCNB. This result robustly substantiates the ability of our system in drastically curtailing the time required for transmission, thereby significantly diminishing the likelihood of data interception, eavesdropping, or tampering with by malicious third parties during transmission. Again, our method not only significantly improves the efficiency of data transmission, but also guarantees the safety and reliability of transmission.

\textit{4) Anti-noise Performance Evaluation} 

To assess the noise reduction capability of our proposed scheme, we consider the MOS metric as shown in Table \ref{table4}. We compare our SSA method with the spectral subtraction and wavelet transform methods in \cite{ref37,ref38,ref39}, to deal with the untreated speech.

\begin{table}[!t]
  \caption{MOS Values for Different Noise Reduction Schemes}
  \label{table4}
  \centering
  \begin{tabular}{|c||c|}
  \hline
  Method & MOS\\
  \hline
  Audio without Noise Reduction & 3.86±0.13 \\
  \hline
  Spectral Subtraction & 3.89±0.17 \\
  \hline
  Wavelet Transform & 3.95±0.14 \\
  \hline
  Ours (SSA) & 4.12±0.23 \\
  \hline
  \end{tabular}
\end{table}

Under the AWGN background, the speech MOS obtained using our SSA method is, 4.12 on average, that is about 0.23 higher than that of the spectral subtraction and 0.17 higher than that of the wavelet transform in Table \ref{table4}. This demonstrates that the SSA-based noise reduction enables our system with a higher clarity and higher quality of user experience, compared to the other two methods.

\textit{5) Complexiy Analysis}

To evaluate the execution efficiency and resource consumption of our proposed algorithm, we select the time complexity and the storage complexity as the metric here. We choose DeepSC-S and DeepSC-SR for comparison. The time complexity comparison is presented in Table \ref{table3}. This result demonstrates that our approach achieves superior algorithmic efficiency, as its time complexity is less than 1\% of that of DeepSC-S and DeepSC-SR while approaching the level of OFI-OFCNB.

\begin{figure}[h!]
  \centering
  \includegraphics[width=\linewidth]{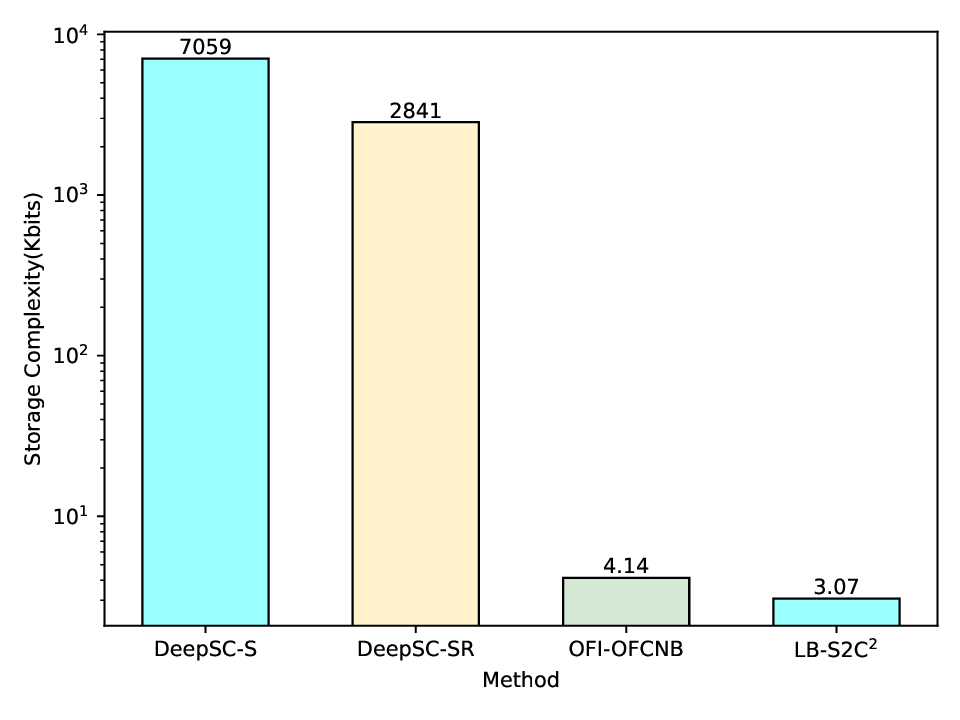}
  \vspace{-0.34in}
  \caption{Storage complexity of different algorithms.}
  \label{fig10}
\end{figure}

We further analyze the storage resource consumption of our algorithm. The required storage space for the key, i.e., the memory occupied to store the encryption key during implementation is considered. 
The comparison  here is based on encrypting the mel spectrogram corresponding to a 3-second audio clip, and recording the memory required to store the keys. The results are shown in Fig. \ref{fig10}.
It can be seen that the storage complexity of our method is approximately 74\% of that of OFI-OFCNB and is significantly smaller than those of DeepSC-S and DeepSC-SR. This indicates that our method possesses a substantial advantage in terms of resource conservation, which is of significantly practical importance in environments with limited storage resources.

\begin{table}[!t]
  \caption{Time Complexity of Different Algorithms}
  \label{table3}
  \centering
  \begin{tabular}{|c||c|}
  \hline
  Method & Time Complexity(FLOPs) \\
  \hline
  DeepSC-S & $24.072 \times 10^9 $ \\ 
  \hline
  DeepSC-SR & $12.401 \times 10^9 $ \\
  \hline
  OFI-OFCNB & $5.735 \times 10^7 $ \\
  \hline
  $ \text{LB-S2C}^{2} $ & $7.908 \times 10^7 $ \\
  \hline
  \end{tabular}
\end{table}

\textit{6) Transmission Security Evaluation}

\begin{figure}[h!]
  \centering
  \includegraphics[width=\linewidth]{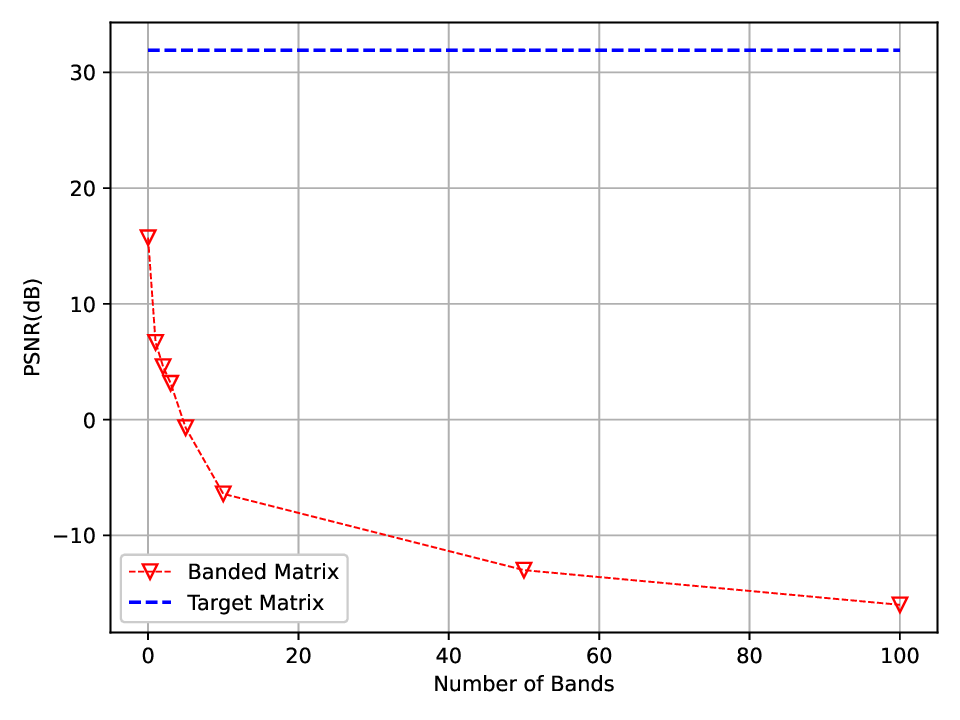}

  \caption{PSNR of the reconstructed spectrum at different band numbers.}
  \label{fig7}
\end{figure}

To verify the security of data transmission, we use the performance metric of PSNR to evaluate the sensitivity of the dictionary matrix. Specifically, since reconstructing the target spectrum requires the dictionary matrix, we will test the sensitivity of our system on the dictionary matrix, and use the PSNR of the reconstructed spectrum serving as the test metric. In our test experiment, for example, the target matrix capable of accurately reconstructing the spectrum is set as the special banded matrix with zero band and all 1s at the diagonal. Whereas, we choose a random-value banded matrix of the same size to reconstruct the spectrum at the receiver.  We set the number of bands to 100 at the starting point and gradually approach 0, recording the PSNR value at each step, which is shown in Fig. \ref{fig7}.

We observe a significant drop in PSNR as the number of bands increases as shown in Fig. \ref{fig7}. This reveals that even only approximately 0.1\% difference  from the target matrix results in a PSNR decrease of more than 20, severely impacting the reconstruction of the target spectrum. And the accurate  reconstruction is feasible only if the dictionary matrix is nearly identical to the one used for the high-order matrix sparsification at the transmitter side. This test underscores the heightened sensitivity of the dictionary matrix used, implying that even minuscule unauthorized modifications during data transmission can be readily detected. Capitalizing on this characteristic, the receiving end can leverage  the PSNR fluctuation monitoring  as an effective means to judge whether the data is tampered with by man-in-the-middle  attacks during transmission. Consequently, this approach serves a safeguard against potential breaches in data integrity and confidentiality.  In essence, this test not only validates the high sensitivity of the utilized dictionary matrix, but also enables a practical method for ensuring the communication security through monitoring  PSNR changes.

Moreover, we can prove that it is impossible for an eavesdropper to decipher the dictionary matrix. The hypothesis of the experiment is that the eavesdropper intercepts the data during transmission and attempts to decrypt it by brute force. In the testing experiment, the dictionary matrix is set as a random banded matrix (with one band, dimensions of 896×896), and the traversal step size is set to 7. Thus, traversing all possible cases would require $256^{128\times128}$ computations. With this computational cost and complexity, it is  highly unlikely that an eavesdropper can use brute force to decipher the dictionary matrix,further verifying the security of the transmission.


\textit{7) Information Privacy Evaluation}

In addition to the transmission security, the role of speaker recognition as a cornerstone of information privacy is also of crucial importance, with its significance in safeguarding personal privacy.     We will compare our $ \text{LB-S2C}^{2} $ with the two semantic speech communication methods(DeepSC-S and DeepSC-SR) and OFI-OFCNB. In the experiment, we construct an audio sample set based on the Aishell dataset to calculate the average voiceprint similarity. Notably, to balance the quality of audio restoration and compression speed, our method operates with a compression ratio of 10\%.

\begin{figure}[h!]
  \centering
  \includegraphics[width=0.98\linewidth]{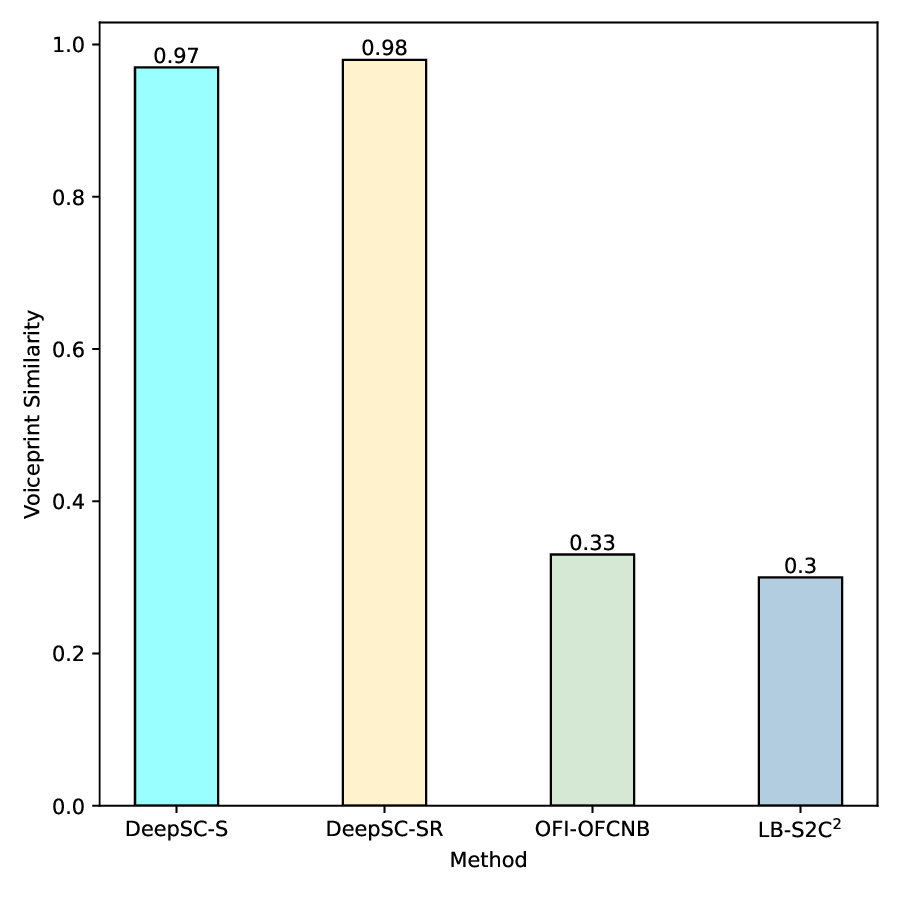}

  \caption{Comparison of voiceprint similarity for different methods.}
  \label{fig8}
\end{figure} 

Fig. \ref{fig8} shows the comparison of voiceprint similarity. It becomes evident that the mean voiceprint similarity of  $ \text{LB-S2C}^{2} $ is a mere 0.30, which is much lower than the similarity values 0.97 and 0.98 of the DeepSC-S and DeepSC-SR systems,and lower than 0.33 of OFI-OFCNB system. This pronounced difference robustly demonstrates that compared to the popular semantic speech communication schemes, our method significantly enhances the difficulty in illicitly obtaining the speaker's voiceprint characteristics, thereby enhancing the protection for the identity information and other privacy aspects.

\begin{table}[!t]
  \caption{Entropy and zero frequency of different methods}
  \label{table2}
  \centering
  \resizebox{1\columnwidth}{!}{
  \begin{tabular}{|c||c|c||c|}
  \hline
  \multirow{2}{*}{\makecell{Method}} & \multicolumn{2}{c||}{Entropy} & \multirow{2}{*}{\makecell{Zero Frequency \\ (Count)}} \\
  \cline{2-3}
  & Obtained Value & Ideal Value & \\
  \hline
  $ \text{LB-S2C}^{2} $ & 6.08 & 9.74 & 0 \\ 
  \hline
  DeepSC-S & 1.13 & 6.74 & 206 \\
  \hline
  DeepSC-SR & 2.46 & 5.38 & 255 \\
  \hline
  OFI-OFCNB & 2.87 & 8.00 & 127 \\
  \hline
  \end{tabular}
  }
\end{table}

More importantly, the entropy and the number of zero frequencies are also considered to further evaluate the information security of our $ \text{LB-S2C}^{2} $ system \cite{ref44,ref45}.\ As shown in Table. \ref{table2}, the entropy of transmission data for $ \text{LB-S2C}^{2} $ is much larger than that of the DeepSC-S, DeepSC-SR and OFI-OFCNB systems, and  the  entropy ratio of the practically obtained entropy to the ideal value is calculated as 62.4\%, which is higher than the ratios of 16.8\% for DeepSC-S, 45.7\% for DeepSC-SR and 35.9\% for OFI-OFCNB.\ Moreover, the number of zero frequencies  is 0 for our $ \text{LB-S2C}^{2} $, whereas the other three have much larger number of zero frequencies.\ All these phenomena imply that our $ \text{LB-S2C}^{2} $ system retains more randomness and unpredictability in the information transmitted, which helps to resist the potential attacks and privacy theft, thus enhancing the  robustness and reliability of our analog cryptographic system. 

\begin{figure}[h!]
  \centering
  \includegraphics[width=\linewidth]{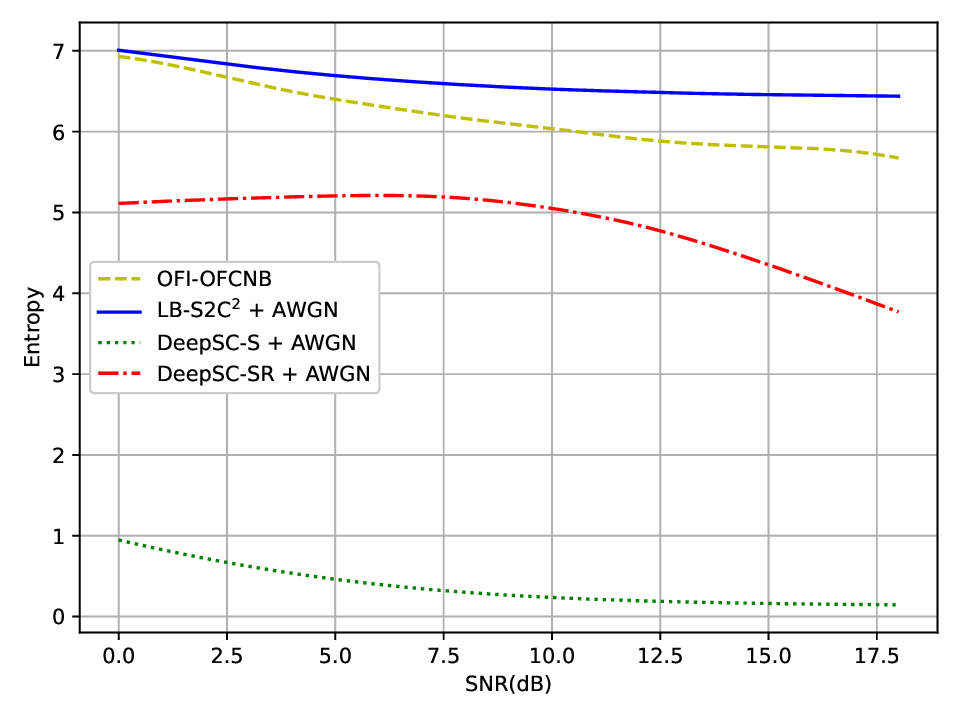}
  \vspace{-0.3in}
  \caption{The practical entropy of different systems in AWGN channel.}
  \label{fig9}
\end{figure}  

Note that Table. \ref{table2} shows the entropy evaluated without transmission noise considered. Next, we test the practical entropy of the system in AWGN channel, with different signal-to-noise ratios (SNRs) to denote the different levels of AWGN.\ As shown in Fig. \ref{fig9}, when the SNR drops from 20dB to 0dB, the practical entropy of $ \text{LB-S2C}^{2} $ remains almost the same, while the practical entropy of DeepSC-S, DeepSC-SR and OFI-OFCNB increases by about 1.3, 0.8 and 1.2, respectively.\ This is due to the higher variance of the transmitted data in $ \text{LB-S2C}^{2} $, its sensitivity to noise is much lower than that of the other two.\ This implies that our $ \text{LB-S2C}^{2} $ system has higher integrity and authenticity than the other systems, due to the more robust anti-noise capability. This can also guarantee the transmission security.

\section{CONCLUSION}

In this paper, we have proposed the secure speech communication system model $ \text{LB-S2C}^{2} $ based on the structured spectral compression. By combining the waveform segmentation with the structured spectrum coding, our model reduces the compression time to be less than 10\% and achieves the compression ratio of around 5\%, significantly improving the compression efficiency. More importantly, it necessitates a coding rate of only 3.9kbps without the privacy disclosure, which is lower than the 6.3 kbps required by G.723 and the 8 kbps required by G.729. It demonstrates a high privacy protection capability, with an average voiceprint similarity of only 0.3, much lower than the popular semantic speech systems. It also has a lower time complexity of only $O(n)$ and a smaller memory requirement of only 12 bits, thereby ensuring effective resource savings without additional overhead. In addition, a noise reduction module is implemented to effectively improve the reconstructed speech quality, resulting in a high MOS value around 4.12. Extensive experimental results have verified that our speech communication system can efficiently achieve the low-bitrate and high-security speech communication for the narrowband IoT-NTN.

Due to the limitation of communication bandwidth, the individual users' speech quality would be degraded with the increasing of users in the narrowband NTN communication. Thus, it is significant to explore the lower-bitrate speech coding scheme without the privacy disclosure in the future work.

\bibliography{bare_jrnl_new_sample4.bib}

\end{document}